\documentclass[12pt]{iopart}

\usepackage{graphicx}
\usepackage{amssymb}
\usepackage{geometry}
\usepackage{xcolor}
\usepackage{hyperref}
\hypersetup{pdfstartview=FitH, linkcolor=linkcolor,citecolor=citecolor,colorlinks=true} 

\begin{document}
\definecolor{linkcolor}{HTML}{008000}
\definecolor{citecolor}{HTML}{4682B4}
\title{Optical-mechanical cooling of a charged resonator}
\author{Dmitry N. Makarov}
\address{Northern (Arctic) Federal University, nab. Severnoi Dviny 17, 163002, Arkhangelsk, Russia.}
\ead{makarovd0608@yandex.ru}
\vspace{10pt}
\begin{indented}
\item[] November 2018
\end{indented}
\begin{abstract}
One of the most effective methods for cooling micro and nano devices to ultra low temperatures is the sideband method. Currently, this approach is being studied experimentally and theoretically. Theoretical results that relate to this method correspond to the case of the interaction of an electromagnetic field with a micro or nano resonator through radiation pressure. Obviously, if you choose a charged micro or nano resonator, the interaction will be different. In this paper, this case is considered and it is shown that cooling will be as effective as in the sideband method. The resulting final equation in the work, for the average value of the quantum number of the cooled resonator, has a surprisingly simple analytical form, which allows for the most complete analysis of the system under study.
\end{abstract}

\section{Introduction}
Currently, there are many ways to optically cool a substance. The most effective method of cooling micro and nano devices is the sideband method \cite{Wilson_2007,Genes_2008,Gigan_2006,SCHLIESSER_2008,Chan_2011,Teufel_2011}. This method shows great promise for using quantum effects (including coherent properties) in hybrid systems \cite{Machnes_2012,Aspelmeyer_2014} but first it is necessary to prepare the mechanical component in the ground state, i.e. cool it as much as possible. The sideband method is based on the connection of a mechanical device, that is, a resonator (target), with a microwave or optical resonator (auxiliary) \cite{Aspelmeyer_2014,Marquardt_2007}. Moreover, the frequency of the microwave or optical resonator should be high enough to be in the ground state at ambient temperature. Note that with modern technologies, the average number of photons in such a resonator can be very small \cite{Tian_2009,Wang_2011,Schmidt_2011,Triana_2015} and in theoretical calculations this value is usually chosen as $ n_{aux} = 0 $. The connection of a mechanical resonator with an electromagnetic field occurs through pressure by radiation. 
Although this interaction is generally not linear, under real conditions the relationship can be considered linear. In theory, such a relationship is given in the form \cite{Aspelmeyer_2014,Marquardt_2007} $ {\hat {H}}_{int} = g(t) q x $, where $ g(t) $ is a communication parameter, $ x $ is a variable mechanical mode and $ q $ is a variable electromagnetic mode. Currently, a system with a Hamiltonian in the form of ${\hat{H}}={\hat{H}}_{t}+{\hat{H}}_{aux}+{\hat{H}}_{int}$ is being actively studied, where $ {\hat{H}}_{t} $ is the Hamiltonian of the target and $ {\hat{H}}_{aux} $ is the Hamiltonian of the electromagnetic field, with the presence of environment and losses in the cavity \cite{Wilson_2007,Teufel_2011,Wang_2011,Schmidt_2011,Triana_2015,Safavi_2013,Palomaki_2013}. The evolution of the density matrix in the presence of an environment is usually described using a master equation, for example, in the model of Ullersma-Caldeira-Leggett \cite{Ullersma_1966,Caldeira_1981}, and many theoretical results have been confirmed experimentally, which confirms the adequacy of the applied model. Currently, the study of such systems is mainly directed towards the optimal selection of the connecting function $ g(t) $ for maximum cooling of the target \cite{Wang_2011,Schmidt_2011,Triana_2015,Frank_2016}. For example, in \cite{Triana_2015} it was shown that with the optimal choice of the function $ g(t) $ taking into account non-Markov evolution processes of the system under consideration, the average value of the quantum number (phonon) of the target can be $n_t=10^{-3}$.

In spite of this, the problem of cooling charged mechanical resonators (charged targets) is relevant. Charged mechanical resonators can be not only simple mechanical systems, but also molecular ions, which expands the field of use of the results obtained in the work. So far, the cooling of such systems has not been implemented, and indeed, such studies are rarely found in the literature.

In this work, it will be shown that the mechanical charged resonator can be cooled to ultra-small quantum states, similar to the method of the sideband. As an example, we will consider the case of cooling a mechanical resonator without taking into account the external environment, where, with certain target parameters and an electromagnetic field, the target can be cooled to the ground quantum state. The main equation showing the average number of quantum states of $ n_t $ charged resonator has a surprisingly simple analytical form, which allows for the most complete analysis of the $ n_t $ quantity under investigation.

\section{The model and its solution}
Consider the interaction of a charged mechanical resonator (charged target) with a quantized electromagnetic field (auxiliary). The Schrodinger equation for such a system will be in the form ${\hat H}\Psi=i\hbar \frac{\partial \Psi}{\partial t}$, where the Hamiltonian ${\hat H}$ is 
\begin{equation}
{\hat H}= \hbar\omega\left({\hat a}^{+} {\hat a}+\frac{1}{2}\right)+\hbar\Omega\left({\hat b}^{+} {\hat b}+\frac{1}{2}\right)+{\hat {H}}_{int},
\label{1}
\end{equation}
where $\omega$ is the frequency of the mechanical mode of the resonator, $ \Omega $ is the frequency of the electromagnetic mode, ${\hat a},{\hat b}$ are operators of annihilation of the mechanical and electromagnetic modes, respectively, and $ {\hat{H}}_{int} $ is the Hamiltonian responsible for the interaction of the electromagnetic field with a charged target. Obviously, in the case of interaction with a target having an electric charge of $ Z $ and a mass of $ M_t $, this interaction will be
\begin{equation}
{\hat {H}}_{int}= \frac{Z}{c M_t}\hat{\bf{ A}}\hat{\bf{ p}}+\frac{1}{2M_t}\left(\frac{Z}{c}\hat{\bf{ A}}\right)^2 ,
\label{2}
\end{equation}
where ${\hat{\bf{ A}}}$ is the vector potential of the electromagnetic field, and $\hat{\bf{ p}}$ is the moment operator of a mechanical resonator. Since a one-dimensional mechanical resonator is considered, $\hat{\bf{ p}}=-i\hbar \frac{\partial}{\partial x} {\bf i}$, where $ {\bf i} $  is the unit vector (mechanical resonator is one-dimensional). Since we are interested in the microwave or optical part of the electromagnetic spectrum, we can apply the dipole approximation, in which the coordinate representation $ {\hat{\bf{A}}} = \sqrt{\frac {4 \pi c^2} {\Omega V}}{\bf u} q $, where $ q $ is the field variable, $ V $ is the volume of space in which the electromagnetic field is located, and $ {\bf u} $ is the polarization of the electromagnetic field. Further, for convenience, we use a system of units, where $ \hbar = 1, M_t = 1, Z = 1 $. As a result, we must consider the Hamiltonian
\begin{equation}
{\hat{H}}= \frac{\Omega}{2}\left(q^2-\frac{\partial^2 }{\partial {q^2}}\right)+\frac{\omega}{2}\left(x^2-\frac{\partial^2 }{\partial {x^2}}\right)+\frac{\beta^2}{2}q^2-i\beta {\bf u i} \sqrt{\omega}q \frac{\partial }{\partial x} ,
\label{3}
\end{equation}
where $\beta=\sqrt{\frac{4\pi}{\Omega V}}$.

A similar Hamiltonian, but set to another problem, was considered in \cite{Makarov_SREP_2018} (see also \cite{Makarov_2017_adf,Makarov_2018_PRE}), where an analytical solution was found for the nonstationary Schrodinger equation with $ \beta \ll 1 $. Indeed, for a realistic microcavity or focal volume \cite{Tey_2008}, $ \beta $ takes values of the order of $ 10^{-5} - 10^{-3} $, and usually it is much less than even these values. We write out the basic equation for our case using the results \cite{Makarov_SREP_2018}.
The wave function of the system under consideration $ \Psi(x,q,t) $ will have the form
\begin{eqnarray}
\Psi(x,q,t)=\sum^{s_1+s_2}_{m_1=0}a_{m_1,s_1+s_2-m_1}(t)\Phi_{m_1}(x)\Phi_{s_1+s_2-m_1}(q), 
\label{4}
\end{eqnarray}
where $ \Phi_{m}(z) $ are known wave functions of a harmonic oscillator in a state with a quantum number $ m $. In the case under consideration, $ s_1, s_2 $ are the quantum numbers of the initial states, respectively, for the charged mechanical resonator (charged target) and the electromagnetic field. As shown in \cite{Makarov_SREP_2018}, in equation (\ref{4}) a law of conservation of quantum numbers applies, $ s_1 + s_2 = m_1 + m_2 $, therefore in (\ref{4}) $ m_2 $ is replaced by $ m_2 = s_1 + s_2-m_1 $. The coefficient $ a_{m_1, s_1 + s_2-m_1}(t) $ is determined by the equation
\begin{eqnarray}
a_{m_1, m_2}(t)=\sum^{s_1+s_2}_{n=0}A^{s_1,s_2}_{n,s_1+s_2-n}A^{*{m_1,m_2}}_{n,s_1+s_2-n}e^{-i{\delta n}t}, 
\label{5}
\end{eqnarray}
where 
\begin{eqnarray}
\delta = \beta \sqrt{\Omega}\left( \alpha + \epsilon \right), ~~\alpha = \sqrt{\frac{\omega}{\Omega}}\left(\epsilon \mp \sqrt{\epsilon^2+1} \right),~~ \epsilon =\frac{\Omega^2-\omega^2 +\beta^2 \Omega}{2\beta \sqrt{\Omega}\omega}.
\label{6}
\end{eqnarray}
In equation (\ref{6}) for $ \alpha $ with $ \epsilon>0 $, the upper sign, must be used; with $ \epsilon<0 $, the lower sign applies. As shown in \cite{Makarov_SREP_2018} with $ \beta \ll 1 $ coefficient $ \alpha \in (-1,1) $ and coefficient
\begin{eqnarray}
A^{s_1,s_2}_{n,m}=\frac{\alpha^{s_{2}+m}\sqrt{n!m!}(-1)^{s_2+m}i^{s_1-n}}{(1+\alpha^2)^{\frac{s_1+s_2}{2}}\sqrt{s_{1}!s_{2}!}}P^{(-(1+s_1+s_2), n-s_2)}_{m}\left(-\frac{2+\alpha^2}{\alpha^2} \right),
\label{7}
\end{eqnarray}
where $ P^{(b, c)}_{a}(x) $ is the Jacobi polynomial.

Thus, the solution of the Schrodinger equation for the problem in question has been found; next, we turn to the results of finding the average number of phonons $ n_t $ in such a system.

\section{Results}

The average number of quantum states (phonons) of a charged resonator will be
\begin{eqnarray} 
n_t=\sum^{s_1+s_2}_{m_1=0}m_1|a_{m_1, s_1+s_2-m_1}(t)|^2. 
\label{7}
\end{eqnarray}
Indeed, the probability amplitude to detect a charged resonator in the state $ |m_1\rangle $ will be $ a_{m_1} = \langle m_1|\Psi(x, q, t)\rangle $, and from (\ref{4}) it is easy to see that $ a_{m_1} =a_{m_1, s_1 + s_2-m_1}(t) $. Equation (\ref{7}) is the basis for further analysis. It should be added that the resulting equation for $ n_t $ depends on two parameters, $ \alpha $ with $ \delta t $, and $ n_t(\alpha,\delta t) = n_t (-\alpha,\delta t) $. In other words, the average number of phonons is an even function with respect to $ \alpha $, which allows us to use only $ \alpha \in (0,1) $ for further analysis.

\begin{figure}[!h]
\center{\includegraphics[angle=0,width=0.7\textwidth, keepaspectratio]{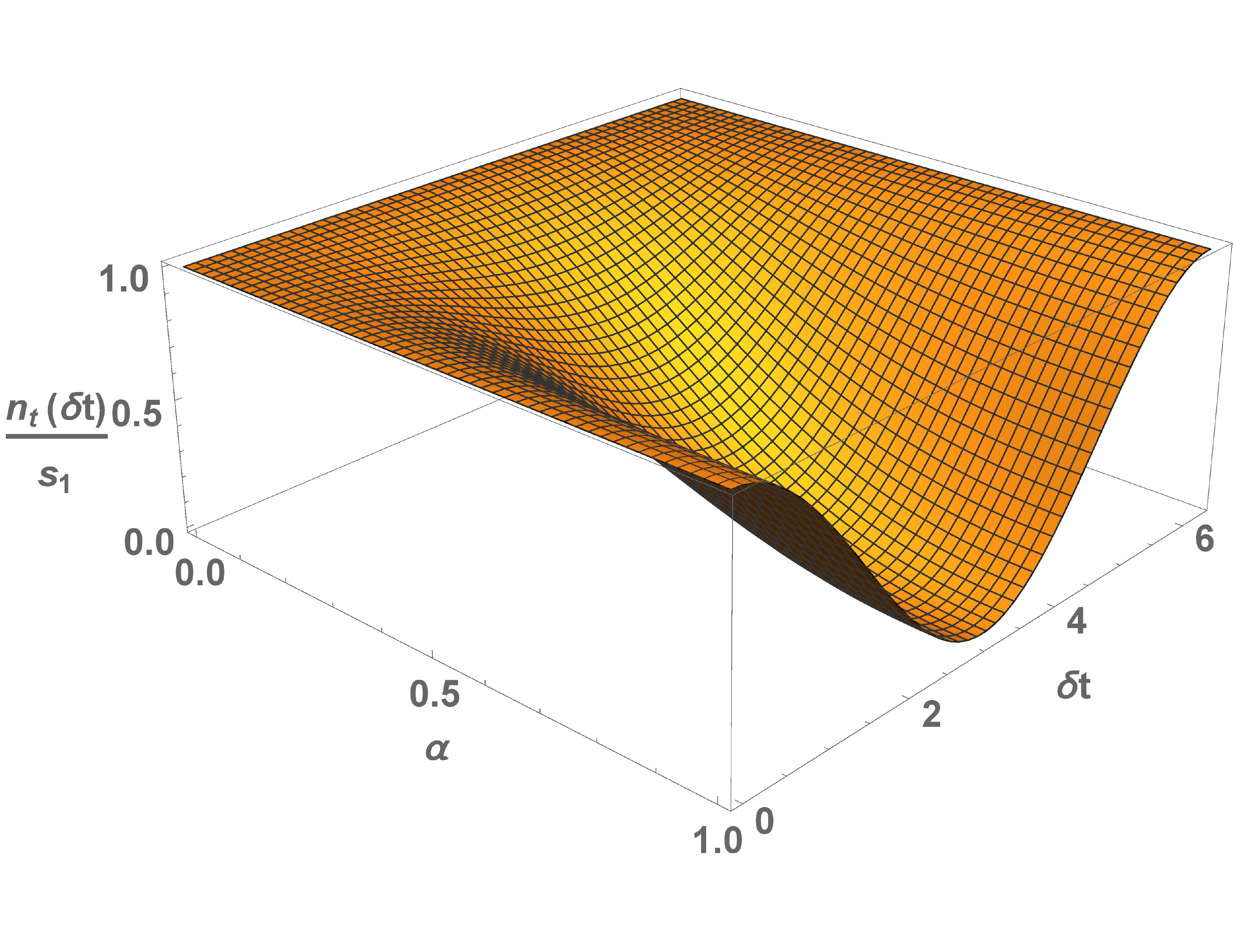}} 
\caption[fig_1]{3D graph of the function $ n_t /s_1 $ as a function of the parameters $ \alpha, \delta t $}
\label{fig_1}
\end{figure}

\begin{figure}[!h]
\begin{minipage}[h]{0.49\linewidth}
\center{\includegraphics[angle=0,width=1\textwidth, keepaspectratio]{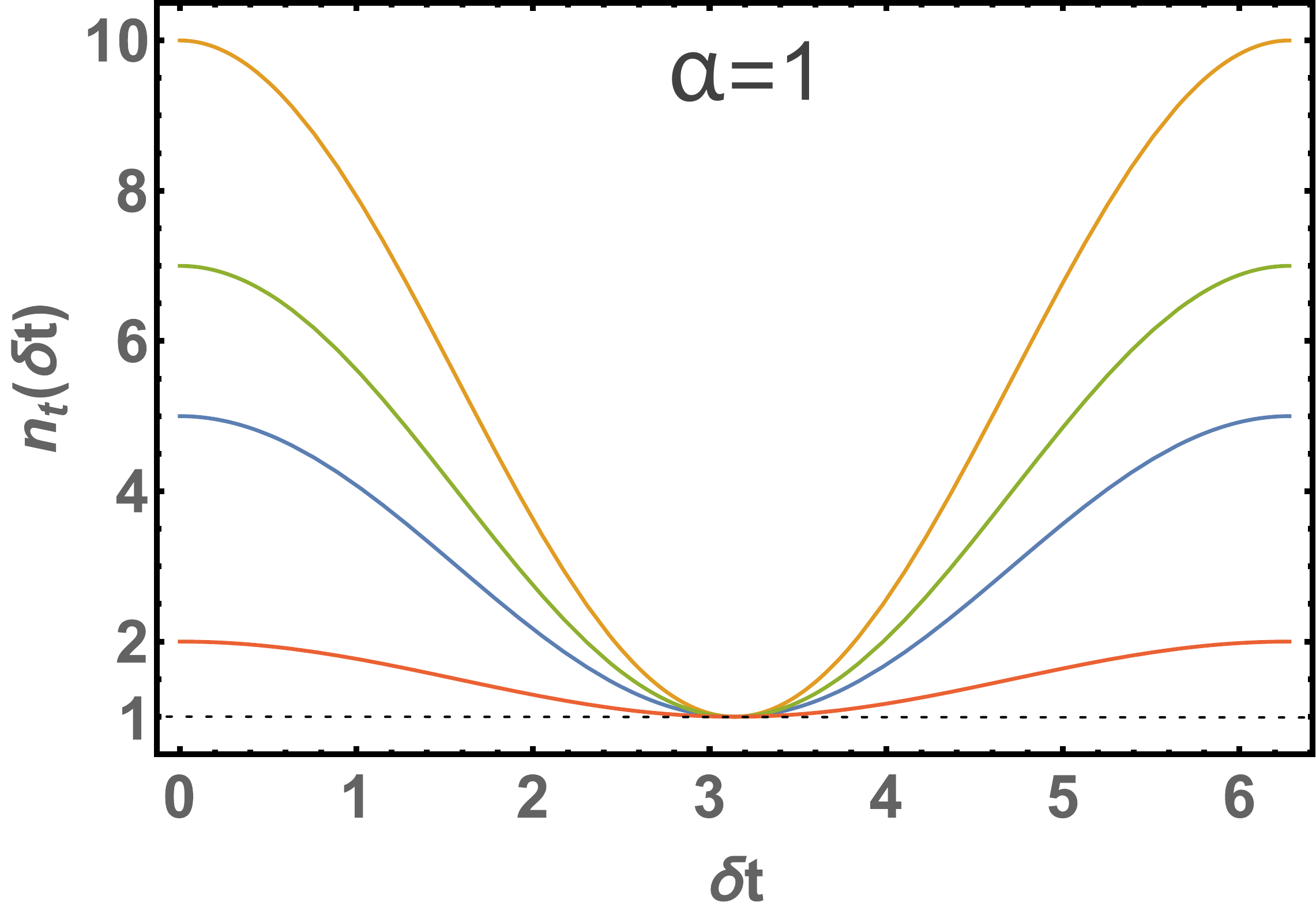}} \\
\end{minipage}
\hfill
\begin{minipage}[h]{0.49\linewidth}
\center{\includegraphics[angle=0,width=1\textwidth, keepaspectratio]{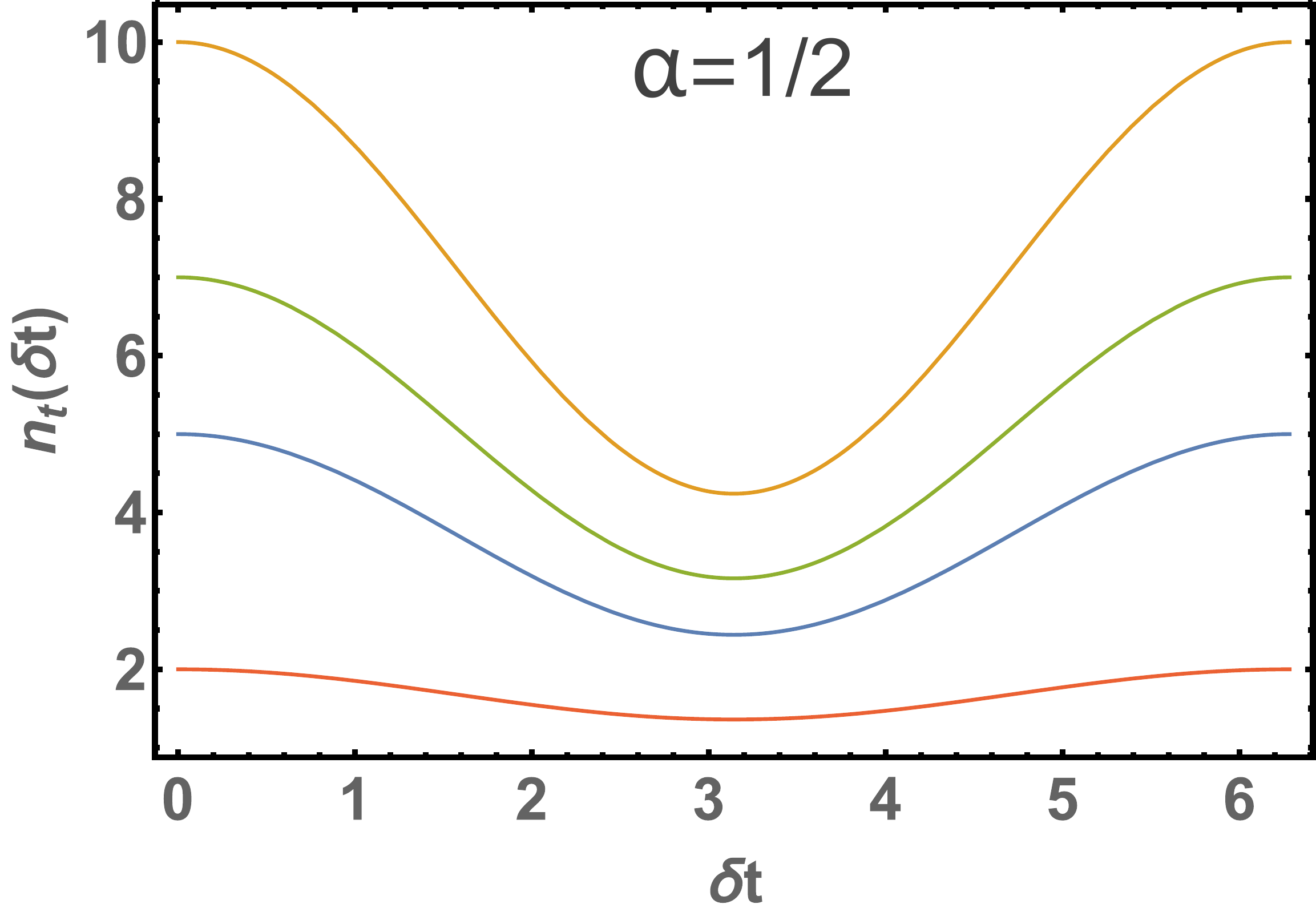}} \\ 
\end{minipage}
\hfill
\begin{minipage}[h]{0.49\linewidth}
\center{\includegraphics[angle=0,width=1\textwidth, keepaspectratio]{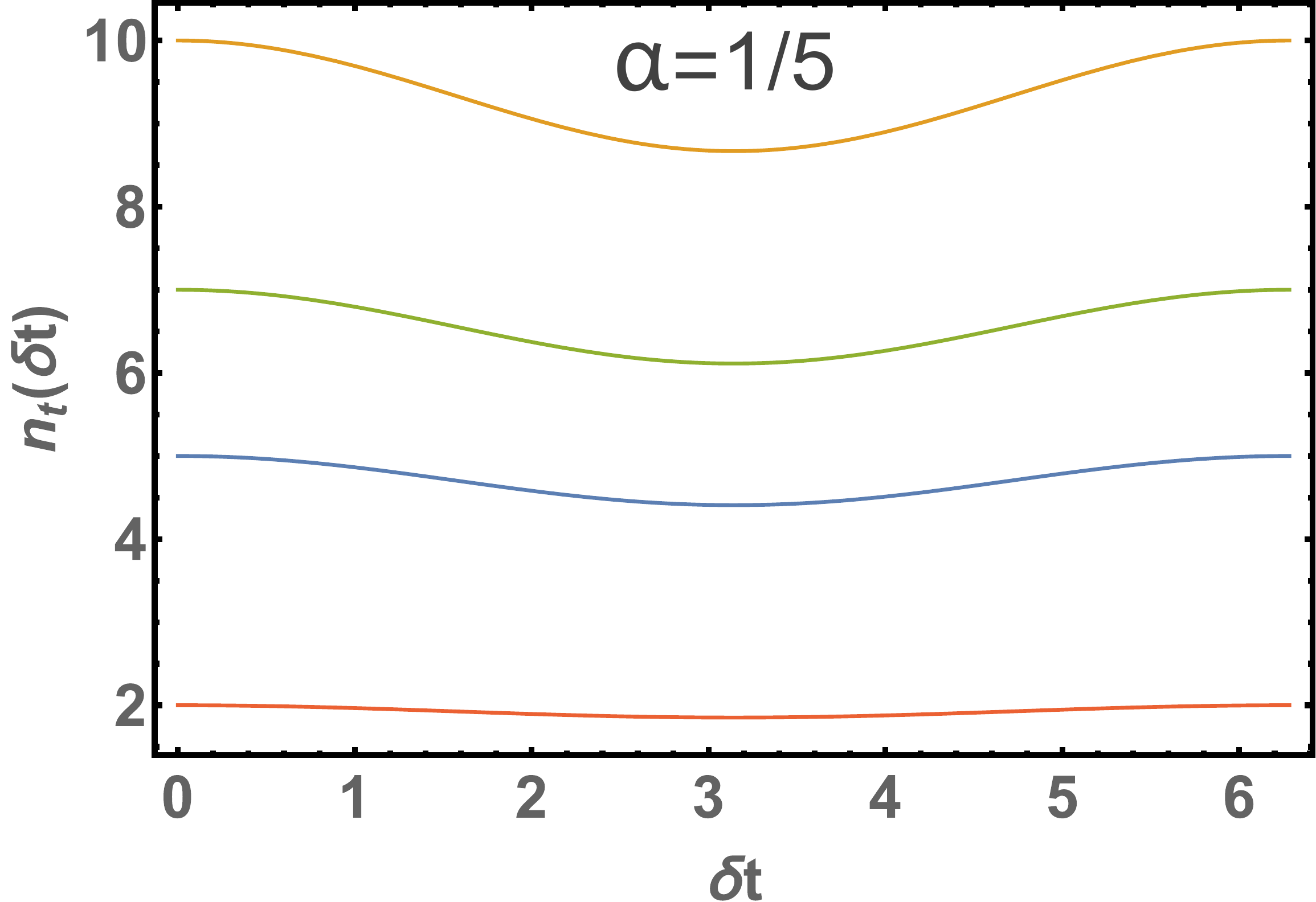}} \\
\end{minipage}
\hfill
\begin{minipage}[h]{0.49\linewidth}
\center{\includegraphics[angle=0,width=1\textwidth, keepaspectratio]{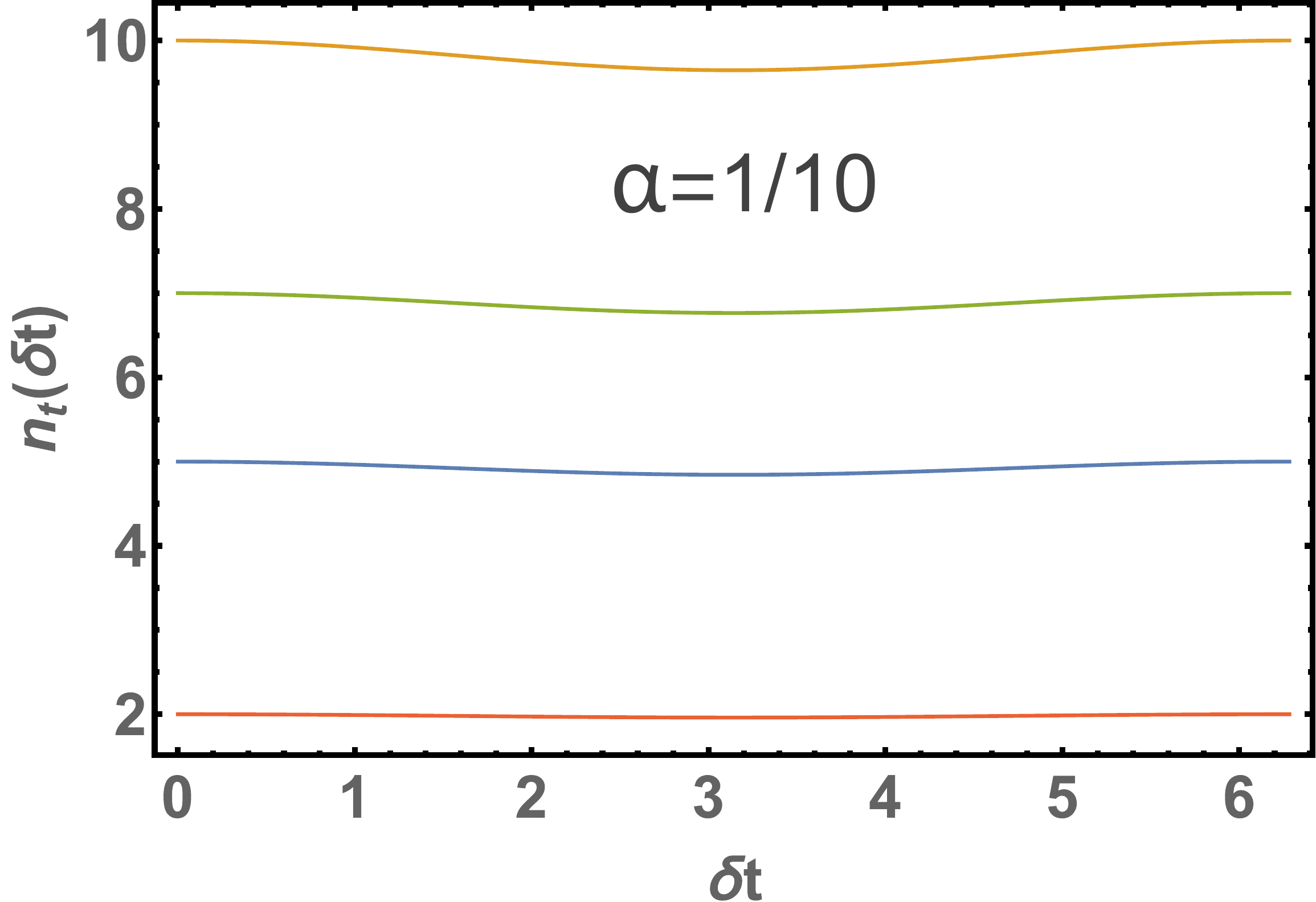}} \\
\end{minipage}
\caption[fig_2]{The dependence is $ n_t = n_t (\delta t) $ when $ s_2 = 1 $, and $ s_1 = (2; 5; 7; 10) $ (on charts it is bottom-up) and a fixed value of $ \alpha $. The dependence on $ s_1 $ on the presented graphs is such that when $ n_t (\delta t = 0) = s_1 $}
\label{fig_2}
\end{figure}

We are interested in the maximum cooling of the charged resonator. As mentioned in the introduction, the number of photons in the resonator, under modern conditions, can be very small and in theoretical calculations one can choose a value tending to zero. This means that in our calculations we choose $ s_2 = 0 $, and $ s_1 $ can be changed arbitrarily, depending on the initial conditions of the problem. If we use the properties of the Jacobi polynomials $ P ^ {(b, c)} _ {a} (x) $, then equation (\ref {7}), for $ s_2 = 0 $, can be calculated analytically, leading to a simple equation
\begin{eqnarray} 
n_t=s_1 \frac{1+\alpha^4+2\alpha^2\cos(\delta t)}{\left( 1+\alpha^2 \right)^2 }. 
\label{8}
\end{eqnarray}
Figure \ref{fig_1} shows the graph of the function $ n_t/s_1 $ depending on the two parameters of the system in question: $ \alpha,\delta t $. It can be seen from figure \ref{fig_1} that for $ \alpha = 1$ and $\delta t=\pi $, the number of phonons $ n_t = 0 $; this result is also easily shown analytically. As an example, we present the results of calculations $n_t = n_t (\alpha, \delta t)$ in figure \ref{fig_2}, with $ s_2 = 1 $, and $ s_1 = (2; 5; 7; 10) $.
Figure \ref{fig_2} shows that the minimum value of $ n_t $ occurs as well as when $ s_2 = 0 $, i.e. with $ \alpha = 1, \delta t = \pi $, but $ min (n_t) = 1 $. Thus, the best cooling of a charged resonator occurs when $ s_2 = 0 $, and $ \alpha = 1, \delta t = \pi $.

Our data show that it is possible to cool the charged resonator to $ n_t = 0 $ (albeit without taking into account various losses) for certain system parameters. We also find the condition for the occurrence of such cooling. As shown, the maximum cooling occurs when $ s_2 = 0 $ and $ \alpha = 1 $; therefore, in further analysis, we assume that the cooling is the most intensive at $ \alpha \approx 1 $. For $ \beta \ll 1 $ and $ \alpha \approx 1 $, based on equation (\ref{6}), the condition $ | \epsilon | \ll 1 $ must be satisfied, and $ \Omega \approx \omega $, and also $ \delta \approx \beta \sqrt{\omega} $. In this case, $ \epsilon = \frac{\Delta \omega}{\beta \sqrt {\omega}} $, where $ \Delta \omega = \Omega- \omega $. As a result, we obtain $ \epsilon \approx \frac{\Delta \omega}{\delta} \ll 1 $, or more simply $ \Delta \omega \ll \delta $.

As a result, the maximum cooling of the resonator will be at $ s_2 = 0 $ when the condition $ \Delta \omega \ll \delta $ is satisfied.

\section{Conclusion}

The results show that there is a theoretical possibility of cooling a charged resonator to the zero quantum state $ n_t = 0 $. It should be added that our results did not take into account any external influences on the system under consideration, which of course would lead to different results. Despite this, the theoretical result reported on here is interesting and suggests that this line of research should be pursued, both theoretically and experimentally. The results obtained in the work have a surprisingly simple analytical form, which is especially true for equation (\ref{8}). The fact that the results were obtained in an analytical form allowed us to single out the fundamentals of the parameters of the system under consideration at which the maximum cooling of the charged resonator occurs. Physically, this means that with these optimal parameters of the system in question, energy exchange between the electromagnetic field and the charged resonator occurs with greatest efficiency, to the fact that the vibrational energy of a charged resonator transforms into the energy of an electromagnetic field, depending on the interaction time. Indeed, it is easy to draw such a conclusion, given that there is a preservation of quantum numbers $ s_1 + s_2 = m_1 + m_2 $ (see above or \cite{Makarov_SREP_2018}) and if we average this expression, we get $ s_1 + s_2 = n_t(\alpha, \delta t) + n_{aux} (\alpha, \delta t) $, where $ n_t $ is the average number of phonons in the system, and $ n_{aux} $ is the average number of photons in the system. Since $ s_1 + s_2 = const $, and $ n_t(\alpha, \delta t) $ has a dependence which is shown in equation (\ref{8}), then reducing the number of phonons by a certain amount increases the number of photons by the same value i.e. there is an exchange of energy. \\

{\bf Funding Information.} Grant of the President of the Russian Federation (No.  MK-6289.2018.2)


\section*{References}

\begin{thebibliography}{1}
\bibitem{Wilson_2007} I. Wilson-Rae, N. Nooshi, W. Zwerger, and T. J. Kippenberg, Theory of Ground State Cooling of a Mechanical Oscillator Using Dynamical Backaction. \emph{Phys. Rev. Lett.} {\bf 99} 093901 (2007). 
\bibitem{Genes_2008}  C. Genes, D. Vitali, et al. Ground-state cooling of a micromechanical oscillator: Comparing cold damping and cavity-assisted cooling schemes, \emph{Phys. Rev. A} {\bf 77} 033804 (2008). 
\bibitem{Gigan_2006} S. Gigan, H. R. Bohm, et al. Self-cooling of a micro-mirror by radiation pressure, \emph{Nature} {\bf 444} 67 (2006).
\bibitem{SCHLIESSER_2008} A. Schliesser, R. Riviere, et al. Resolved-sideband cooling of a micromechanical oscillator, \emph{Nature Physics} {\bf 4} 415-419 (2008). 
\bibitem{Chan_2011} J. Chan, T. P. M. Alegre, et al. Laser cooling of a nanomechanical oscillator into its quantum ground state, \emph{Nature} {\bf 478} 89 (2011).
\bibitem{Teufel_2011} J. D. Teufel, T. Donner, et al. Sideband cooling of micromechanical motion to the quantum ground state.\emph{Nature} {\bf 475} 359 (2011).
\bibitem{Machnes_2012} S. Machnes, J. Cerrillo, et al. Pulsed Laser Cooling for Cavity Optomechanical Resonators. \emph{Phys. Rev. Lett.} {\bf 108} 153601 (2012). 
\bibitem{Aspelmeyer_2014} M. Aspelmeyer, T. J. Kippenberg, and F. Marquardt, Cavity optomechanics. \emph{Rev. Mod. Phys.} {\bf 86} 1391 (2014).
\bibitem{Marquardt_2007} F. Marquardt, J. P. Chen, A. A. Clerk, and S. M. Girvin, Quantum Theory of Cavity-Assisted Sideband Cooling of Mechanical Motion. \emph{Phys. Rev. Lett.} {\bf 99} 093902 (2007).
\bibitem{Tian_2009} L. Tian, Ground state cooling of a nanomechanical resonator via parametric linear coupling. \emph{Phys. Rev. B} {\bf 79} 193407 (2009).
\bibitem{Wang_2011} Xiaoting Wang, Sai Vinjanampathy, Frederick W. Strauch and Kurt Jacobs, Ultraefficient Cooling of Resonators: Beating Sideband Cooling with Quantum Control. \emph{Phys. Rev. Lett.} {\bf 107} 177204 (2011). 
\bibitem{Schmidt_2011} R. Schmidt, A. Negretti, et al. Optimal Control of Open Quantum Systems: Cooperative Effects of Driving and Dissipation. \emph{Phys. Rev. Lett.} {\bf 107} 130404 (2011).
\bibitem{Triana_2015} Johan F. Triana, Andres F. Estrada, and Leonardo A. Pachon, Ultrafast Optimal Sideband Cooling under Non-Markovian Evolution. \emph{Phys. Rev. Lett.} {\bf 116} 183602 (2015).
\bibitem{Safavi_2013} A. H. Safavi-Naeini, J. Chan, et al. M. Aspelmeyer, and O. Painter, Squeezed light from a silicon micromechanical resonator. \emph{New Journal of Physics} {\bf 15} 035007 (2013).
\bibitem{Palomaki_2013} T. A. Palomaki, J. W. Harlow, et al. Coherent state transfer between itinerant microwave fields and a mechanical oscillator. \emph{Nature} {\bf 495} 210-214  (2013).
\bibitem{Ullersma_1966} P. Ullersma, An exactly solvable model for Brownian motion: IV. Susceptibility and Nyquist's theorem. \emph{Physica} {\bf 32} 27 (1966).
\bibitem{Caldeira_1981} A. O. Caldeira and A. J. Leggett, Influence of Dissipation on Quantum Tunneling in Macroscopic Systems. \emph{Phys. Rev. Lett.} {\bf 46} 211 (1981).
\bibitem{Frank_2016} S. van Frank, M. Bonneau, et al. Optimal control of complex atomic quantum systems. \emph{Scientific Reports} {\bf 478} 34187  (2016).
\bibitem{Makarov_SREP_2018} Dmitry N. Makarov, Quantum entanglement of a harmonic oscillator with an electromagnetic feld. \emph{Scientific Reports} {\bf 8} 8204  (2018).
\bibitem{Makarov_2017_adf} D. N. Makarov, High Intensity Generation of Entangled Photons in a Two-Mode Electromagnetic Field. \emph{Annalen der Physik} {\bf 529} 1600408 (2017).
\bibitem{Makarov_2018_PRE} Dmitry N. Makarov, Coupled harmonic oscillators and their quantum entanglement. \emph{Phys. Rev. E} {\bf 97} 042203 (2018).
\bibitem{Tey_2008} M. K. Tey, Z. Chen, et al. Strong interaction between light and a single trapped atom without the need for a cavity. \emph{Nature Physics} {\bf 4} 924 - 927 (2008).
\end{thebibliography}
\end{document}